\documentclass[12pt,preprint]{aastex}
\usepackage{apjfonts}
\usepackage{graphicx}
\usepackage{lineno}
\usepackage{ulem}

\begin{document}

\title{KIC 10975348: A double-mode or triple-mode high-amplitude $\delta$ Scuti star?  }
\shorttitle{$Kepler$ Observations of KIC 10975348}

\author{Tao-Zhi Yang\altaffilmark{1},
Xiao-Ya Sun\altaffilmark{1},
Zhao-Yu Zuo\altaffilmark{1, *},
Hai-Wen Liu\altaffilmark{2}
}

\altaffiltext{1}{
School of Physics, Xi'an Jiaotong University, 710049 Xi'an, PR China; e-mail:zuozyu@xjtu.edu.cn;}

\altaffiltext{2}{
School of Information and Communications Engineering, Xi'an Jiaotong University, 710049 Xi'an, PR China;  }


\begin{abstract}

In this paper, we analyze the light variations of KIC 10975348 using photometric data delivered from $Kepler$ mission. This star is exceptionally faint ($K_{p}$ = 18.6 mag), compared to most well-studied $\delta$ Scuti stars. The Fourier analysis of the short cadence data (i.e. Q14, Q15 and Q16, spanning 220 days) reveals the variations are dominated by the strongest mode with frequency F0 = 10.231899 $\rm{d^{-1}}$, which is compatible with that obtained from $RATS-Kepler$. The other two independent modes with F1 (= 13.4988 $\rm{d^{-1}}$) and F2 (= 19.0002 $\rm{d^{-1}}$) are newly detected and have amplitudes two orders of magnitude smaller than F0. We note that, for the first time, this star is identified to be a high-amplitude $\delta$ Sct (HADS) star with amplitude of about 0.7 mag, and the lower ratio of F0/F1 = 0.758 suggests it might be a metal-rich variable star. The frequency F2 may be a third overtone mode, suggesting this target might be a new radial triple-mode HADS star. We perform $O - C$ analysis using 1018 newly determined times of maximum light and derive an ephemeris formula: $T_{max}$ = 2456170.241912(0)+0.097734(1) $\times$ $E$. The $O - C$ diagram shows that the pulsation period of KIC 10975348 seems to show no obvious change, which is in contrast to that of the majority of HADS stars. The possible cause of that may be due to the current short time span of observations. To verify its possible period variations, regular observation from space with a longer time span in future is needed.

\end{abstract}


\section{Introduction}

The main scientific goal of the $Kepler$ $Mission$ was to search for Earth-like planets outside the solar system by detecting transits of the host star \citep{koch2010,Borucki2010}. The high-precision photometric data delivered from $Kepler$ telescope also provides an unprecedented opportunity to probe into stellar interiors using their natural oscillation modes, and hence greatly expands the field of asteroseismology \citep{{Chaplin2010}}. Thanks to the ultra-high precision photometric observations at the level of $\mu$mag, our understanding of many types of pulsating stars has been significantly improved. For instance, \citet{Bedding2011} reported that in red giants the hydrogen- and helium-burning stars can be distinguished according to the observed period spacings of gravity modes; \citet{2018Natur.554...73G} found a pulsating white dwarf has a large oxygen-dominated core, which exceeds the predictions of existing models of stellar evolution. As a common group of variable stars, $\delta$ Scuti stars are excellent target for asteroseismic  study. At present, more than 2000 $\delta$ Scuti stars have been found in the $Kepler$ field \citep{Balona2011,Balona2014,Bowman2016}, however, only several high-amplitude $\delta$ Scuti stars are found and investigated in detail so far \citep{Balona2012a,Yang2018b,Yang2019}.  

High-amplitude $\delta$ Sct (HADS) stars, as a subgroup of $\delta$ Sct stars, are usually recognized by their relatively simple, non-sinusoidal light variations with peak-to-peak amplitude larger than 0.3 mag \citep{Breger2000}. They are slow rotators with $v$ $sin$ $i$ $<$ 30 km s$^{-1}$ and the slow rotation might be a precondition for their high amplitudes \citep{Breger2000}. In the H-R diagram, HADS stars seem to concentrate in a narrow strip in the $\delta$ Sct instability region with a width of about 300 K in temperature \citep{McNamara2000}, but observations from space photometry revealed that some HADS stars can also be found beyond the narrow region \citep{2016MNRAS.459.1097B}. The light variations of HADS stars are usually dominated by the fundamental and/or first overtone radial mode(s) \citep{Breger2000,McNamara2000}. Although the AAVSO International Variable Star Index (VSX) \citep{Watson2015} lists almost 600 HADS stars, radial triple-mode HADS stars are particular rare. Only five HADS stars with three consecutive radial modes, i.e. AC And \citep{1976ApJ...203..616F}, V823 Cas \citep{2006A&A...445..617J}, V829 Aql \citep{1998IBVS.4549....1H}, GSC 762-110 \citep{2008A&A...478..865W} and GSC 03144-595 \citep{2016AJ....152...17M}, are found at present.

In recent decades, nonradial modes with low amplitude are also detected in some HADS stars, owing to extensive high-precision photometric observations \citep{2011A&A...528A.147P}. Moreover, with the advent of space asteroseismology, long-term variations of the principal modes and more low-amplitude frequencies are detected in HADS stars. \citet{Balona2012a} reported a slight amplitude variation of the dominant modes in the HADS star V2367 Cyg and also found this star rotates with a twice the projected rotational velocity of any other HADS star. With $Kepler$ observations, amplitude modulation in several HADS stars was investigated. \citet{Yang2018b} found a pair of low-amplitude triplet structures in the frequency spectra of KIC 5950759 and the reason for this triplet structure might be the amplitude modulation of stellar rotation. Another HADS star KIC 10284901 also shows a weak amplitude modulation with two frequencies, and analysis suggests that they might be related to the Blazhko effect \citep{Yang2019}. Hence, the identification of these low-amplitude frequencies in HADS stars possesses great potential to improve our understanding of the stellar interior, and the comparison of single-, double- and triple-mode HADS stars may illuminate what determines the number of radial modes in which a star pulsates. More HADS stars with high-precision photometric observations are needed.

KIC 10975348 ($\alpha_{2000}$=$19^{h}$$26^{m}$$46^{s}$.1, $\delta_{2000}$=+$48^{\circ}$$25^{'}$$30^{''}$.8, 2MASS: J1926461+4825303) was classified as a $\delta$ Sct star with a pulsation period of 2.35 hrs by \cite{{2014MNRAS.437..132R}}. In that paper, KIC 10975348 was reported as a mid A type star based on spectrum from Gran Telescopio Canarias (GTC). Some basic parameters of this target were listed in Table \ref{tab:basic_parmeters}. According to the amplitude of the light curves, KIC 10975348 appears to be a HADS star, yet the exact type is uncertain. In this work, using the high-precision photometric data from $Kepler$ mission, we further verify its nature, and investigate its long-term periodic variation as well. 

\begin{deluxetable}{lcc} 
\tabletypesize{\small} 
\tablewidth{0pc} 
\tablenum{1} 
\tablecaption{Basic Properties of KIC 10975348 \label{tab:basic_parmeters}} 
\tablehead{ 
\colhead{Parameters}    & 
\colhead{KIC 10975348}  &
\colhead{References}       
}
\startdata 
   $K_P$     &  18.598 mag &  a  \\
   $P$       &  2.35 h     &  b  \\
   $i$       &  18.496 mag &  a  \\
   $g$       &  18.89 mag  &  b  \\
   $U-g$     &   0.19 mag  &  b  \\
   $g-r$     &   0.34 mag  &  b  \\
   \enddata 
   
   \tablecomments{(a) These parameters are available in the KASOC: https://kasoc.phys.au.dk/; (b) \citealt{2014MNRAS.437..132R}. }
\end{deluxetable}

\section{OBSERVATIONS AND DATA REDUCTION}

\begin{figure*}
\begin{center}
  \includegraphics[width=0.95\textwidth,trim=45 10 55 20,clip]{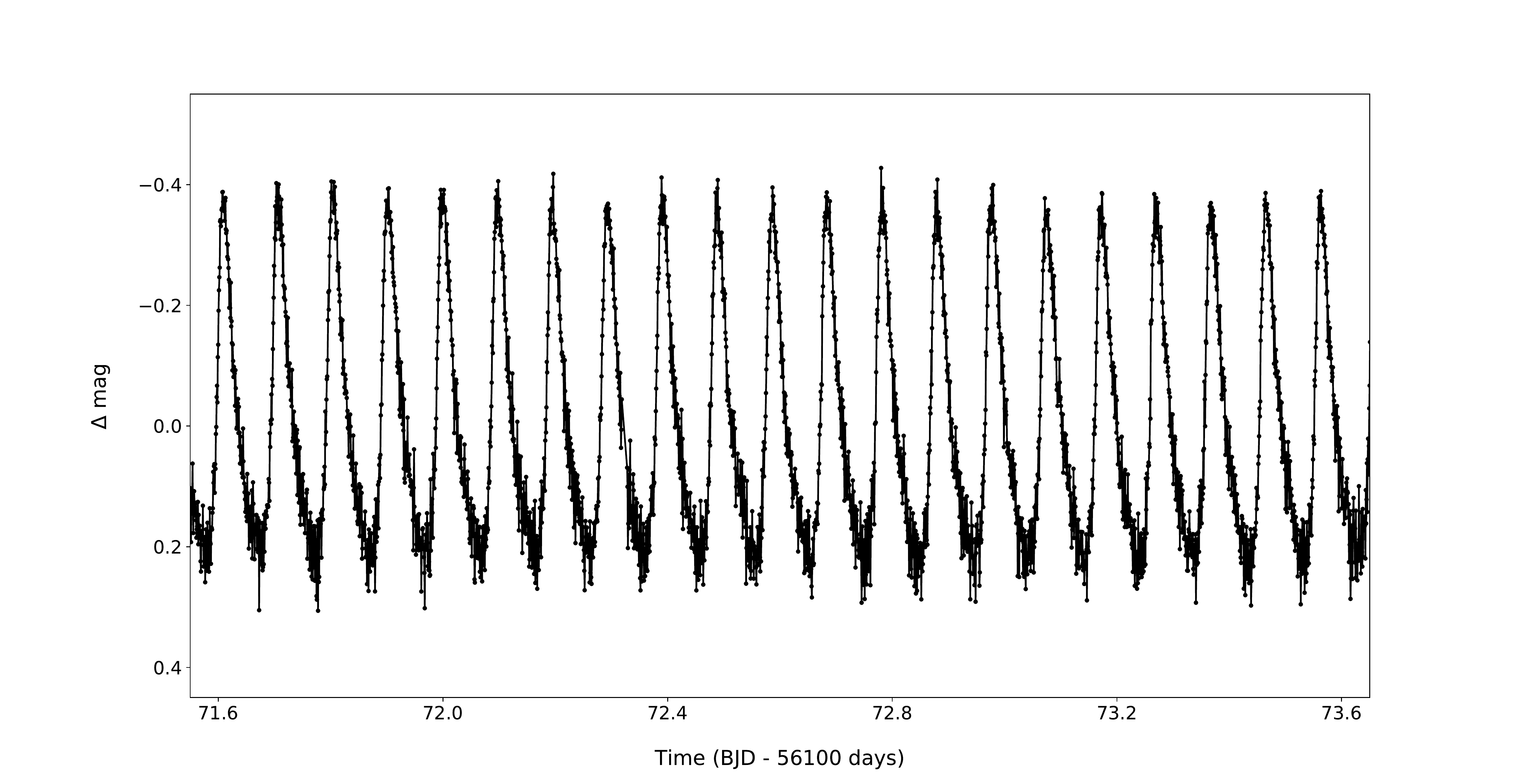}
  \caption{A portion of the short cadence light curve of KIC 10975438. The amplitude of the light curve is about 0.7 mag, revealing that it may be a HADS star.}
    \label{fig:light curve}
\end{center}
\end{figure*}

KIC 10975348 was observed by $Kepler$ space telescope from BJD 2456107.139 to 2456424.001, including three quarters (i.e. Q14, Q15 and Q16). There are two observation strategies for this star: Long cadence (LC: 29.5-min integration time) mode and Short cadence (SC: 58.5-s integration time). To avoid the Nyquist alias peaks, we only used the SC data, i.e. Q14.3, Q15.3 and Q16.3 in this work. More details on how to identify the Nyquist alias peaks in LC data can be found in \citet{2013MNRAS.430.2986M} and \cite{Yang2018b}. All the SC data are available in Kepler Asteroseismic Science Operations Center (KASOC) data base\footnote{KASOC data base: {http://kasoc.phys.au.dk}}\citep{Kjeldsen2010}, in which two types of data: "raw" flux and "corrected" flux, are provided. The former data has been reduced by the NASA $Kepler$ Science pipeline in fact, and the "corrected" one can be obtained from KASOC Working Group 4 (WG\#4: $\delta$ Sct stars). The corrected flux was used in this work and we first performed several corrections to the data including removing the obvious outliers and de-trending the light curve with a low-order polynomial. Then the flux data was converted to the magnitude scale, and each quarter was adjusted to zero by subtracting their mean value. The final rectified light curve was obtained with 147481 data points in total, spanning over about 220 days. Figure \ref{fig:light curve} shows a portion of the rectified light curve of KIC 10975348 covering 2 days. From this figure, it is clear that the amplitude of the light curve is about 0.7 mag, which is larger than the typical amplitude ($>$ 0.3 mag) of high-amplitude $\delta$ Sct star.

\section{FREQUENCY ANALYSIS}

We used the software PERIOD04 \citep{Lenz2005} to analyze the pulsating behavior of KIC 10975348. The rectified light curve was fitted with the following formula:      
\begin{equation}
m = m_{0} + \Sigma\mathnormal{A}_{i}sin(2\pi(\mathnormal{f}_{i}\mathnormal{t} + \phi_{i})), \label{equation1}
\end{equation}
where $m_{0}$, $A_{i}$, $f_{i}$, and $\phi_{i}$ are zero-point, amplitude, frequency and the corresponding phase, respectively. 

In order to detect more significant frequencies, we chose a frequency range of 0 $<$ $\nu$ $<$ 80 d$^{-1}$, which is wider than the typical period range of $\delta$ Scuti stars. During the extraction of significant frequency, the highest peak was usually selected as a potential significant frequency. Then a multi-frequency least-square fit using formula 1 was applied to the light curve with all significant frequencies detected, and obtained the solutions for all the frequencies. A theoretical light curve constructed by the above solutions was subtracted from the rectified data and the residual was obtained for next search. The above steps were repeated until there was no significant peak in the frequency spectrum. The criterion of S/N > 4.0 suggested by \cite{Breger1993} was adopted to judge the significance of the detected peaks. The uncertainties of frequencies were obtained following the method provided by \citet{1999DSSN...13...28M}.

\begin{figure*}
\begin{center}
  \includegraphics[width=0.95\textwidth,trim=45 15 60 40,clip]{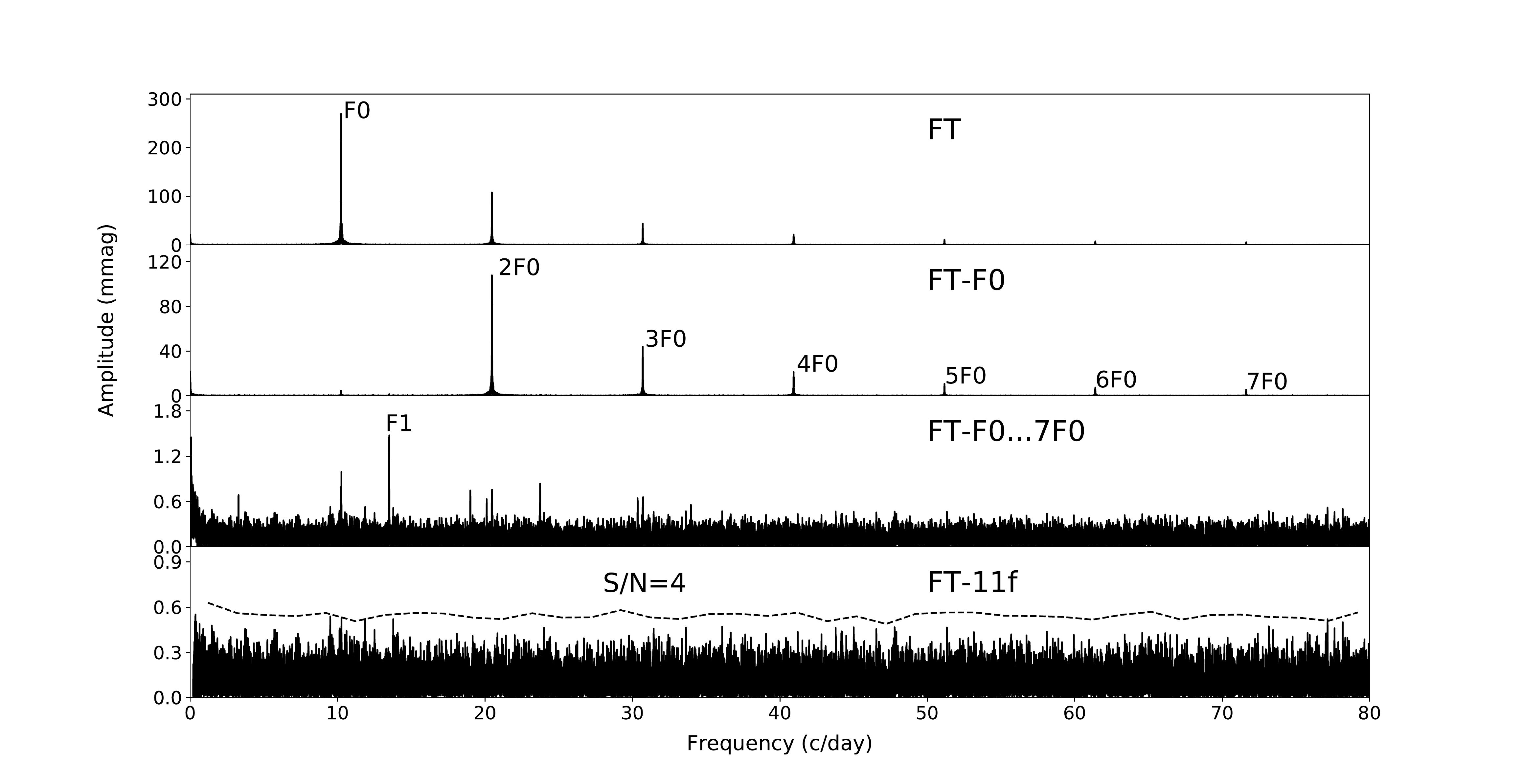}
  \caption{Fourier amplitude spectra and the pre-whitening process for the light curve of KIC 10975348. The first two panels show F0 and its 6 harmonic frequencies. The third panel is the amplitude spectra after subtracting F0 and its harmonic frequencies, where the independent frequency F1 is marked. The bottom panel shows the residual after subtracting 11 significant frequencies. The dotted curve refers to detection limit of S/N = 4.0. }
    \label{fig:amplitude spectra}
\end{center}
\end{figure*}

Figure \ref{fig:amplitude spectra} shows the amplitude spectra and the prewhitening procedures of the light curve. The last panel shows amplitude spectra after 11 detected frequencies were pre-whitened. Following \cite{Breger1993}, we draw the significance curves at signal-to-noise ratio S/N = 4 in last panel for the judgement of a significant peak. There is no significant peak in the residuals and the overall distribution of the residual is typical of noise.
 
A total of 11 significant frequencies were detected in this work and a full list was given in Table \ref{tab:Frequency-SC}. Among these frequencies, three stronger frequencies were considered to be independent. It is reasonable that the strongest peak $f_{S1}$ was assumed to be the fundamental mode, since the light variations were dominated by this frequency. Therefore, we marked $f_{S1}$ with 'F0' in the last column of Table \ref{tab:Frequency-SC}. In addition, six harmonics of F0 were also detected and listed in the table. The other two independent frequencies are $f_{S8}$ and $f_{S10}$, we marked them with 'F1' and 'F2', respectively. The rest of the frequencies are combinations with F0. 

For frequencies F0 and F1, we found the period ratio of $P_{1} / P_{0}$ = F0 / F1 = 0.758 was in the typical range (0.756 - 0.787) of period ratio of the first overtone and fundamental mode for $\delta$ Scuti stars \citep{{Petersen1996}}. It seems that F1 can be identified as the first overtone mode. Considering its peak-to-peak amplitude over 0.7 mag from the light curve, KIC 10975348 seems to be classified as a new double-mode HADS star.  

\begin{deluxetable}{cccccc} 
\tabletypesize{\small} 
\tablewidth{0pc} 
\tablenum{2} 
\tablecaption{A full list of 11 significant frequencies detected in this work (denoted by $f_{Si}$).\label{tab:Frequency-SC}} 
\tablehead{ 
\colhead{$f_{Si}$}   & 
\colhead{Frequency (d$^{-1}$)}  &
\colhead{Amplitude (mmag)}      &
\colhead{S/N}            &
\colhead{Identification} &  
}
\startdata 
    1  &  10.231899(1)  &  269.1   & 984.6 & F0     &  \\
    2  &  20.463798(3)  &  107.9   & 634.7 & 2F0    &  \\
    3  &  30.695470(7)  &   43.9   & 270.8 & 3F0    &  \\
    4  &  40.92731(2)  &   21.5   & 151.4 & 4F0    &  \\
    5  &  51.15921(3)  &   10.8   &  75.7 & 5F0    &  \\
    6  &  61.39110(4)  &    7.5   &  57.0 & 6F0    &  \\
    7  &  71.62278(6)  &    5.6   &  39.6 & 7F0    &  \\
    8  &  13.4988(2)  &    1.5   &   9.6 & F1     &  \\
    9  &  23.7314(4)  &    0.8   &   5.8 & F0+F1  &  \\
    10 &  19.0002(5)  &    0.7   &   5.4 & F2     &  \\
    11 &  33.9629(6)  &    0.5   &   4.1 & 2F0+F1 &  \\   
   \enddata 
    \tablecomments{Among these frequencies, 3 peaks are independent frequencies, others are harmonic or combinations.}
\end{deluxetable}  

\section{$O - C$ Diagram}

To examine the potential long-term period changes of KIC 10975348, a classical $O - C$ diagram was constructed. $O$ is the observed time of maximum light and $C$ is the theoretical value from ephemeris formula \citep{{Sterken2005}}. For the calculations of $O$, we first visually inspected the light curve and determined the preliminary times of maximum, and then made a second-order polynomial to fit the part of light curve around one-third of the full amplitude. It is reasonable to do that since this part of light curve is almost symmetric for this star. The times of extrememum of the polynomial fit were considered as the observed times of maximum light. The typical uncertainties (about 0.00008 d$^{-1}$) were estimated by Monte Carlo simulations. In total, 1018 times of maximum light were obtained and a full list of each quarter is given in Table \ref{tab:max 143}, \ref{tab:max 153} and \ref{tab:max 163}, respectively.

To obtain the calculated times of maximum light, we used all the observed times of maximum and derived a new ephemeris formula:
 
\begin{equation}
T_{max} = T_{0}+P \times E = 2456170.241912(0)+0.097734(1) \times E.
\end{equation}

where $T_{0}$ is the initial epoch, $P$ is the period, and $E$ is the cycle number.

Then the values of $O - C$ were derived based on this new ephemeris. All the $O - C$ values, as well as the corresponding cycle numbers are listed in Table \ref{tab:max 143}, \ref{tab:max 153} and \ref{tab:max 163}, respectively.

\begin{figure*}
\begin{center}
  \includegraphics[width=0.95\textwidth,trim=85 280 90 280,clip]{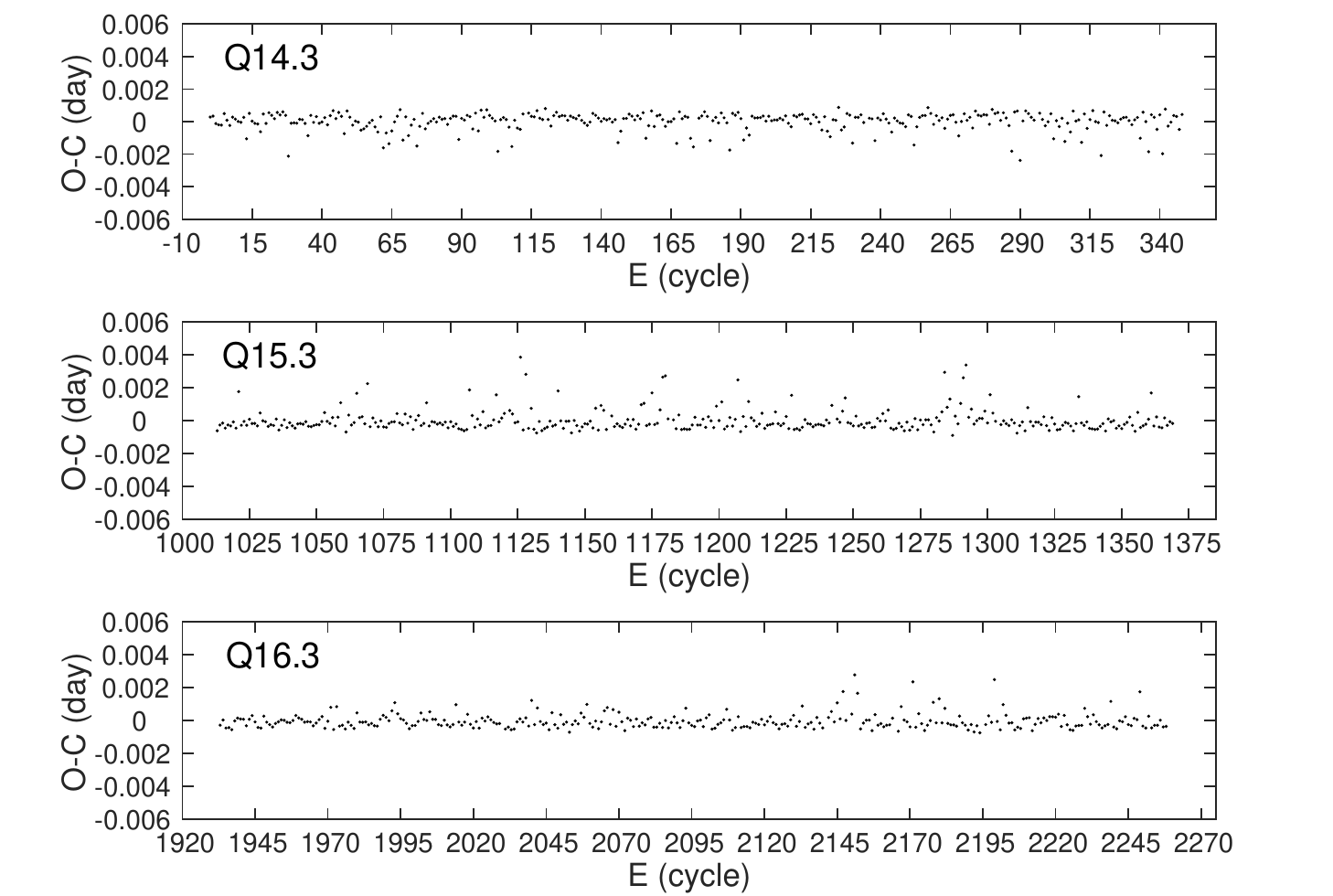}
  \caption{$O-C$ diagram of KIC 10975348, based on a collection of 1018 times of maximum light. Upper panel: The $O-C$ of Q14.3 changes over time. Middle panel: The $O-C$ of Q15.3 changes over time. Lower panel: The $O-C$ of Q16.3 changes over time. None of these panels shows a significant trend, indicating that the pulsation period has no significant change.}
    \label{fig:o-c}
\end{center}
\end{figure*}

Figure \ref{fig:o-c} shows the $O - C$ diagram of three quarters used in this work. It is noteworthy that the $O - C$ of each quarter is almost flat, which suggests the star seems to have no obvious period change. 

\section{DISCUSSION}

\subsection{Double-mode or multi-mode?}

HADS stars are usually single- or double-mode radial pulsators \citep{Breger2000}. As mentioned above, in KIC 10975348, the first two stronger frequencies F0 and F1 give a period ratio of P1/P0 = 0.758, suggesting this star appears to be a double-mode HADS star. A detailed diagram about the double-mode HADS stars and metallicities were given by \cite{Petersen1996}. Their study showed that higher values of the period ratio can be found in metal-poor stars. For KIC 10975348, the lower period ratio of 0.758 implies it might be a metal-rich double-mode HADS star.

The third independent frequency F2 (=19.0002 d$^{-1}$) is interesting. Identification of this frequency is very important, as the pulsating stars with three radial modes are very rare at present \citep{2008A&A...478..865W,2016AJ....152...17M}. \citet{{Stellingwerf1979}} calculated a series of stellar modes for $\delta$ Scuti stars and presented the period ratios of the first four radial modes as: 
$P_{1}$/$P_{0}$ = (0.756 - 787), $P_{2}$/$P_{0}$ = (0.611 - 632) and $P_{3}$/$P_{0}$ = (0.500 - 525), in which $P_{0}$, $P_{1}$, $P_{2}$ and $P_{3}$ present the fundamental mode, first overtone, second overtone and third overtone, respectively. For KIC 10975348, the ratio of F0/F2 (= 0.539) is between $P_{2}$/$P_{0}$ and $P_{3}$/$P_{0}$, but close to $P_{3}$/$P_{0}$. It seems to indicate F2 might be a third radial overtone. To determine the exact nature of F2, detailed seismic modelling and multi-color photometric observations are still required. We note such models about the period ratios are still lacking \citep{2017ApJ...849...38L,2020MNRAS.499.3034D,2020MNRAS.498.1700R}. We suggest more studies of models (linear or nonlinear) investigating expected period ratios and effects on input physics for HADS stars should be undertaken. If confirmed, KIC 10975348 would be a new radial triple-mode HADS star, and hence enrich the sample of triple-mode variables.

From ground-based observations, some HADS stars are the so-called mono-period pulsating variables, e.g. YZ Boo \citep{{Yang2018a}}, XX Cyg \citep{{Yang2012}}, etc. They usually pulsate with a fast rising from minimum to maximum light and a slow declining, and their light amplitudes are nearly constant. The light curve of KIC 10975348 is similar to that of these stars, but it was identified as a double-mode HADS star, since a low-amplitude first overtone was detected in this star owning to the high-precision photometric observations from space. For instance, in the Fourier amplitude spectra of YZ Boo, the typical amplitudes of noise of the residual are 1.5 mmag, while in KIC 10975348, the second independent frequency F1 has an amplitude of only 1.5 mmag, so that such a weak amplitude will certainly fall in the noise if the observations are obtained from the ground and with relatively short time-series observations. This naturally raises such questions: Will the current so-called mono-period HADS stars become double- or multi- mode stars in the space era? Is there any real mono-period HADS star? What about the relation between the mono- and double-mode HADS stars and their stellar parameters? These questions are of importance for the study of HADS stars, particularly for their seismic modellings. The all-sky $TESS$ observations of HADS stars could provide a well-timed opportunity to address these questions and improve our knowledge of HADS stars.   

\subsection{$O - C$}

The period changes due to stellar evolution for stars in and across the lower part of the classical instability strip allow an observational test of stellar evolution theory, assuming that other physical reasons for period changes can be excluded \citep{Breger2000}. From a theoretical point of view, an evolutionary change in $T_{eff}$ and $M_{bol}$ leads to a period change of size \citep[equation 9 of][]{Breger2000}
\begin{equation}
\frac{1}{P}\frac{dP}{dt}=-0.69\frac{dM_{bol}}{dt}-\frac{3}{T_{eff}}\frac{dT_{eff}}{dt}+\frac{1}{Q}\frac{dQ}{dt}
\end{equation}
where $P$ is the period of a radial pulsation mode in unit of days, $Q$ is the pulsation constant. For a specific mode, the Q value is essential constant for all $\delta$ Scuti stars, hence the term $(1/Q)/(dQ/dt)$ is negligible as it is a very small quantity \citep{Breger2000}. The above relation is then reduced to as follows,
\begin{equation}
\frac{1}{P}\frac{dP}{dt}=-0.69\frac{dM_{bol}}{dt}-\frac{3}{T_{eff}}\frac{dT_{eff}}{dt}
\end{equation}

In the lower instability strip where the $\delta$ Scuti stars are found, stellar evolution leads to increasing periods in most of stars, with predicted increase period changes from 10$^{-10}$ yr$^{-1}$ for the stars on the main sequence to 10$^{-7}$ yr$^{-1}$ for the longer-period evolved variable stars \citep{{1998AA...332..958B}}. Such period changes are observable and also have been observed in some radial $\delta$ Scuti pulsators. \citet{{1998AA...332..958B}} calculated the theoretical periods of the radial fundamental modes and their changes during late main-sequence and post-main-sequence evolution of the 1.8 $M_{\odot}$ model and found the observations are consistent with the predicted values. \citet{{2018ApJ...861...96X}} investigated a HADS star VX Hya with the observed period change from $O - C$ and the stellar evolutionary models, and found the period change can be successfully interpreted by the evolutionary effect.

In pulsating stars, when a pulsation period changes linearly with time, the $O-C$ diagram will present a parabolic form that rises or falls depending on whether the period is increasing or decreasing \citep{{Sterken2005}}. 
For HADS stars, the $O - C$ diagram is a powerful tool to investigate their period changes. 
According to \cite{Breger2000}, HADS stars can be divided into two groups depending on whether the period change is increasing or decreasing. For instance, some HADS stars have an increasing period, i.e., YZ Boo \citep{Yang2018a}; XX Cyg \citep{Yang2012}; GP And \citep{zhou2011}, etc., while others pulsate with a decreasing period, such as: BS Aqr \citep{2011Ap&SS.333..125B}; BE Lyn \citep{2011Ap&SS.333..125B}; DY Peg \citep{Blake2003}, etc. Different values of period changes may suggest stars are in different stages of evolution.

For KIC 10975348 in this work, it was also expected to exhibit an increasing or decreasing period. However, from $O-C$ diagram in Figure \ref{fig:o-c}, it seems that the shape of $O-C$ is nearly flat and does not show any curved part, which is in contrast to the predictions by \cite{Breger2000}. The possible cause for that might be due to the shorter time span of the observations for this star. To verify its period variations, theoretical seismic modelling and regular observations from space with a longer time span in the future are necessary. 

\section{SUMMARY}

We analyzed the pulsating behavior of KIC 10975348 using high-precision photometric observations from $Kepler$ mission, and detected 11 significant frequencies from SC data. Among these frequencies, three independent frequencies, i.e. F0 = 10.231899 d$^{-1}$, F1 = 13.4988 d$^{-1}$ and F2 = 19.0002 d$^{-1}$ were found. The lower period ratio (=0.758) of the first two stronger frequencies (i.e. F0 and F1) suggests that KIC 10975348 might be a metal-rich double-mode HADS star. The third independent frequency might be a third overtone mode. If confirmed, KIC 10975348 would be a new radial triple-mode HADS star, and hence enrich the rare sample of pulsating stars with three radial modes. We also compared KIC 10975348 with current mono-period HADS stars, and highlighted the potential of $TESS$ mission for the study of HADS stars.

The $O - C$ diagram was constructed with 1018 times of maximum light and yielded a new ephemeris epoch: $T_{max}$=$T_{0}+P \times E$=2456170.241912(0)+0.097734(1)$\times$ $E$. The $O - C$ analysis indicated that the period of KIC 10975348 shows no obvious change, which is very unusual. The cause of that might be due to the shorter time span of current observations. To verify the evolutionary state of KIC 10975348, regular observations from space in the future are necessary.

\acknowledgements

We thank the referee for his/her comments that help clarify the paper. This research is supported by the National Natural Science Foundation of China (grant Nos. 11573021, U1938104, 12003020) and the Fundamental Research Funds for the Central Universities. We would like to thank the $Kepler$ science team for providing such excellent data.

\begin{deluxetable}{ccrccrccr}
\tabletypesize{\small}
\tablewidth{0pc}
\tablenum{3}
\tablecaption{345 times of Maximum Light and $O - C$ Values of Q14.3.
\label{tab:max 143}} 
\tablehead{
\colhead{$BJD$}   &
\colhead{$E$}  &
\colhead{$O-C$ }      &
\colhead{$BJD$}            &
\colhead{$E$} &
\colhead{$O-C$ }      &
\colhead{$BJD$}            &
\colhead{$E$} &
\colhead{$O-C$} \\
\colhead{(2400000+)}   &
\colhead{}  &
\colhead{(day)}      &
\colhead{(2400000+)}            &
\colhead{} &
\colhead{ (day)}      &
\colhead{(2400000+)}            &
\colhead{} &
\colhead{(day)}   
}
\startdata
        56170.242260  & 0  & 0.000278  & 56181.579400  & 116  & 0.000291  & 56192.916480  & 232  & 0.000256  \\ 
        56170.340060  & 1  & 0.000340  & 56181.677510  & 117  & 0.000671  & 56193.014380  & 233  & 0.000416  \\ 
        56170.437330  & 2  & $-$ 0.000121  & 56181.774780  & 118  & 0.000206  & 56193.111640  & 234  & $-$ 0.000058  \\ 
        56170.534990  & 3  & $-$ 0.000196  & 56181.872430  & 119  & 0.000127  & 56193.307370  & 236  & 0.000210  \\ 
        56170.632700  & 4  & $-$ 0.000220  & 56181.970840  & 120  & 0.000802  & 56193.405070  & 237  & 0.000169  \\ 
        56170.731140  & 5  & 0.000492  & 56182.067900  & 121  & 0.000131  & 56193.501460  & 238  & $-$ 0.001173  \\ 
        56170.828470  & 6  & 0.000082  & 56182.165230  & 122  & $-$ 0.000273  & 56193.600580  & 239  & 0.000219  \\ 
        56170.925880  & 7  & $-$ 0.000235  & 56182.263610  & 123  & 0.000369  & 56193.698200  & 240  & 0.000099  \\
        56171.024120  & 8  & 0.000271  & 56182.361550  & 124  & 0.000571  & 56193.796320  & 241  & 0.000486  \\ 
        56171.121740  & 9  & 0.000150  & 56182.459020  & 125  & 0.000309  & 56193.893720  & 242  & 0.000153  \\ 
        56171.219320  & 10  & 0.000004  & 56182.556820  & 126  & 0.000377  & 56193.991210  & 243  & $-$ 0.000090  \\ 
        56171.317010  & 11  & $-$ 0.000042  & 56182.654530  & 127  & 0.000352  & 56194.088390  & 244  & $-$ 0.000641  \\ 
        56171.415050  & 12  & 0.000262  & 56182.751990  & 128  & 0.000081  & 56194.186980  & 245  & 0.000216  \\ 
        56171.511470  & 13  & $-$ 0.001055  & 56182.849980  & 129  & 0.000335  & 56194.284540  & 246  & 0.000041  \\
        56171.610760  & 14  & 0.000507  & 56182.947630  & 130  & 0.000248  & 56194.382140  & 247  & $-$ 0.000091  \\ 
        56171.708020  & 15  & 0.000026  & 56183.045290  & 131  & 0.000173  & 56194.479850  & 248  & $-$ 0.000116  \\ 
        56171.805610  & 16  & $-$ 0.000110  & 56183.143220  & 132  & 0.000370  & 56194.577390  & 249  & $-$ 0.000312  \\ 
        56171.903310  & 17  & $-$ 0.000151  & 56183.240650  & 133  & 0.000075  & 56194.675850  & 250  & 0.000412  \\ 
        56172.000570  & 18  & $-$ 0.000622  & 56183.338230  & 134  & $-$ 0.000082  & 56194.773480  & 251  & 0.000314  \\ 
        56172.099380  & 19  & 0.000460  & 56183.435810  & 135  & $-$ 0.000233  & 56194.869460  & 252  & $-$ 0.001440  \\ 
        56172.196550  & 20  & $-$ 0.000113  & 56183.533750  & 136  & $-$ 0.000029  & 56194.968320  & 253  & $-$ 0.000321  \\ 
        56172.294940  & 21  & 0.000547  & 56183.632020  & 137  & 0.000501  & 56195.066630  & 254  & 0.000261  \\ 
        56172.392510  & 22  & 0.000388  & 56183.729630  & 138  & 0.000384  & 56195.164470  & 255  & 0.000368  \\ 
        56172.490040  & 23  & 0.000176  & 56183.827200  & 139  & 0.000215  & 56195.262210  & 256  & 0.000375  \\ 
        56172.588150  & 24  & 0.000560  & 56183.924730  & 140  & 0.000013  & 56195.360430  & 257  & 0.000857  \\ 
        56172.685740  & 25  & 0.000410  & 56184.022630  & 141  & 0.000179  & 56195.457810  & 258  & 0.000499  \\ 
        56172.783660  & 26  & 0.000599  & 56184.120290  & 142  & 0.000102  & 56195.554870  & 259  & $-$ 0.000169  \\ 
        56172.881180  & 27  & 0.000383  & 56184.218090  & 143  & 0.000176  & 56195.653140  & 260  & 0.000363  \\ 
        56172.976400  & 28  & $-$ 0.002126  & 56184.315620  & 144  & $-$ 0.000029  & 56195.750570  & 261  & 0.000064  \\ 
        56173.076160  & 29  & $-$ 0.000098  & 56184.413510  & 145  & 0.000121  & 56195.848480  & 262  & 0.000235  \\ 
        56173.173920  & 30  & $-$ 0.000079  & 56184.509830  & 146  & $-$ 0.001293  & 56195.945560  & 263  & $-$ 0.000414  \\ 
        56173.271630  & 31  & $-$ 0.000099  & 56184.608260  & 147  & $-$ 0.000591  & 56196.043860  & 264  & 0.000151  \\ 
        56173.369600  & 32  & 0.000139  & 56184.706770  & 148  & 0.000187  & 56196.141820  & 265  & 0.000378  \\ 
        56173.467320  & 33  & 0.000118  & 56184.804520  & 149  & 0.000196  & 56196.239600  & 266  & 0.000422  \\ 
        56173.564820  & 34  & $-$ 0.000112  & 56184.902510  & 150  & 0.000454  & 56196.336870  & 267  & $-$ 0.000039  \\ 
        56173.661800  & 35  & $-$ 0.000865  & 56185.000120  & 151  & 0.000333  & 56196.433770  & 268  & $-$ 0.000877  \\ 
        56173.760780  & 36  & 0.000383  & 56185.097630  & 152  & 0.000104  & 56196.532390  & 269  & 0.000006  \\ 
        56173.858080  & 37  & $-$ 0.000051  & 56185.195610  & 153  & 0.000357  & 56196.630590  & 270  & 0.000481  \\ 
        56173.956170  & 38  & 0.000303  & 56185.293170  & 154  & 0.000179  & 56196.727920  & 271  & 0.000076  \\ 
        56174.053490  & 39  & $-$ 0.000106  & 56185.391270  & 155  & 0.000544  & 56196.825850  & 272  & 0.000266  \\ 
        56174.151310  & 40  & $-$ 0.000021  & 56185.487430  & 156  & $-$ 0.001027  & 56196.922920  & 273  & $-$ 0.000392  \\
        56174.249310  & 41  & 0.000241  & 56185.585940  & 157  & $-$ 0.000254  & 56197.021700  & 274  & 0.000647  \\ 
        56174.346610  & 42  & $-$ 0.000194  & 56185.684580  & 158  & 0.000655  & 56197.119180  & 275  & 0.000396  \\ 
        56174.444900  & 43  & 0.000365  & 56185.781360  & 159  & $-$ 0.000301  & 56197.216860  & 276  & 0.000343  \\ 
        56174.542940  & 44  & 0.000672  & 56185.879740  & 160  & 0.000350  & 56197.314680  & 277  & 0.000426  \\ 
        56174.640190  & 45  & 0.000185  & 56185.977530  & 161  & 0.000408  & 56197.412390  & 278  & 0.000402  \\ 
        56174.738300  & 46  & 0.000560  & 56186.074980  & 162  & 0.000120  & 56197.509550  & 279  & $-$ 0.000165  \\ 
        56174.835800  & 47  & 0.000325  & 56186.172300  & 163  & $-$ 0.000293  & 56197.608190  & 280  & 0.000743  \\ 
        56174.932460  & 48  & $-$ 0.000742  & 56186.270320  & 164  & $-$ 0.000007  & 56197.705680  & 281  & 0.000495  \\ 
        56175.031590  & 49  & 0.000653  & 56186.368060  & 165  & $-$ 0.000004  & 56197.803480  & 282  & 0.000557  \\ 
        56175.128940  & 50  & 0.000271  & 56186.465960  & 166  & 0.000168  & 56197.900730  & 283  & 0.000075  \\ 
        56175.226190  & 51  & $-$ 0.000219  & 56186.562190  & 167  & $-$ 0.001343  & 56197.998930  & 284  & 0.000540  \\ 
        56175.324170  & 52  & 0.000030  & 56186.661870  & 168  & 0.000604  & 56198.096340  & 285  & 0.000220  \\ 
        56175.421840  & 53  & $-$ 0.000037  & 56186.759170  & 169  & 0.000175  & 56198.193610  & 286  & $-$ 0.000247  \\ 
        56175.519080  & 54  & $-$ 0.000525  & 56186.857060  & 170  & 0.000324  & 56198.289770  & 287  & $-$ 0.001821  \\ 
        56175.616900  & 55  & $-$ 0.000444  & 56186.954730  & 171  & 0.000263  & 56198.389910  & 288  & 0.000590  \\ 
        56175.714790  & 56  & $-$ 0.000288  & 56187.051170  & 172  & $-$ 0.001029  & 56198.487710  & 289  & 0.000655  \\ 
        56175.812740  & 57  & $-$ 0.000067  & 56187.148370  & 173  & $-$ 0.001559  & 56198.582400  & 290  & $-$ 0.002386  \\ 
        56175.910620  & 58  & 0.000078  & 56187.345610  & 175  & 0.000207  & 56198.682560  & 291  & 0.000041  \\ 
        56176.007950  & 59  & $-$ 0.000328  & 56187.443480  & 176  & 0.000346  & 56198.780920  & 292  & 0.000660  \\ 
        56176.105370  & 60  & $-$ 0.000640  & 56187.541470  & 177  & 0.000602  & 56198.878490  & 293  & 0.000495  \\ 
        56176.204030  & 61  & 0.000289  & 56187.638920  & 178  & 0.000319  & 56198.975990  & 294  & 0.000268  \\ 
        56176.299860  & 62  & $-$ 0.001614  & 56187.735170  & 179  & $-$ 0.001167  & 56199.073540  & 295  & 0.000086  \\ 
        56176.398510  & 63  & $-$ 0.000701  & 56187.834130  & 180  & 0.000057  & 56199.269440  & 297  & 0.000517  \\ 
        56176.495590  & 64  & $-$ 0.001359  & 56187.932060  & 181  & 0.000259  & 56199.366790  & 298  & 0.000131  \\ 
        56176.594100  & 65  & $-$ 0.000576  & 56188.029460  & 182  & $-$ 0.000078  & 56199.464140  & 299  & $-$ 0.000256  \\ 
        56176.692390  & 66  & $-$ 0.000020  & 56188.127800  & 183  & 0.000527  & 56199.562120  & 300  & $-$ 0.000012  \\ 
        56176.790480  & 67  & 0.000332  & 56188.225240  & 184  & 0.000233  & 56199.660340  & 301  & 0.000478  \\ 
        56176.888620  & 68  & 0.000734  & 56188.322660  & 185  & $-$ 0.000075  & 56199.756540  & 302  & $-$ 0.001059  \\ 
        56176.984470  & 69  & $-$ 0.001141  & 56188.418730  & 186  & $-$ 0.001747  & 56199.855380  & 303  & 0.000055  \\ 
        56177.083640  & 70  & 0.000291  & 56188.518710  & 187  & 0.000505  & 56199.952760  & 304  & $-$ 0.000306  \\ 
        56177.180220  & 71  & $-$ 0.000860  & 56188.616330  & 188  & 0.000394  & 56200.051000  & 305  & 0.000202  \\ 
        56177.278600  & 72  & $-$ 0.000221  & 56188.714240  & 189  & 0.000566  & 56200.147310  & 306  & $-$ 0.001223  \\ 
        56177.376690  & 73  & 0.000135  & 56188.811590  & 190  & 0.000182  & 56200.246170  & 307  & $-$ 0.000096  \\ 
        56177.472780  & 74  & $-$ 0.001502  & 56188.908010  & 191  & $-$ 0.001130  & 56200.343360  & 308  & $-$ 0.000638  \\ 
        56177.571730  & 75  & $-$ 0.000290  & 56189.006480  & 192  & $-$ 0.000394  & 56200.442380  & 309  & 0.000648  \\ 
        56177.670260  & 76  & 0.000505  & 56189.103780  & 193  & $-$ 0.000833  & 56200.539500  & 310  & 0.000037  \\ 
        56177.767040  & 77  & $-$ 0.000447  & 56189.202680  & 194  & 0.000334  & 56200.637310  & 311  & 0.000108  \\ 
        56177.865110  & 78  & $-$ 0.000108  & 56189.300300  & 195  & 0.000226  & 56200.733660  & 312  & $-$ 0.001275  \\ 
        56177.962970  & 79  & 0.000021  & 56189.398040  & 196  & 0.000231  & 56200.833130  & 313  & 0.000466  \\ 
        56178.060870  & 80  & 0.000179  & 56189.495790  & 197  & 0.000244  & 56200.929980  & 314  & $-$ 0.000420  \\ 
        56178.158400  & 81  & $-$ 0.000026  & 56189.593620  & 198  & 0.000341  & 56201.028260  & 315  & 0.000124  \\ 
        56178.256420  & 82  & 0.000267  & 56189.691040  & 199  & 0.000031  & 56201.126550  & 316  & 0.000677  \\ 
        56178.354030  & 83  & 0.000139  & 56189.789080  & 200  & 0.000330  & 56201.223600  & 317  & $-$ 0.000002  \\ 
        56178.451810  & 84  & 0.000185  & 56189.886530  & 201  & 0.000053  & 56201.321280  & 318  & $-$ 0.000058  \\
        56178.549180  & 85  & $-$ 0.000172  & 56189.984320  & 202  & 0.000104  & 56201.416990  & 319  & $-$ 0.002085  \\ 
        56178.647280  & 86  & 0.000194  & 56190.082040  & 203  & 0.000096  & 56201.517390  & 320  & 0.000590  \\ 
        56178.745160  & 87  & 0.000335  & 56190.179890  & 204  & 0.000209  & 56201.614310  & 321  & $-$ 0.000231  \\ 
        56178.842890  & 88  & 0.000328  & 56190.277520  & 205  & 0.000108  & 56201.712630  & 322  & 0.000362  \\ 
        56178.939190  & 89  & $-$ 0.001099  & 56190.375500  & 206  & 0.000351  & 56201.810090  & 323  & 0.000081  \\ 
        56179.038220  & 90  & 0.000194  & 56190.472780  & 207  & $-$ 0.000104  & 56201.907790  & 324  & 0.000049  \\ 
        56179.135860  & 91  & 0.000097  & 56190.570300  & 208  & $-$ 0.000317  & 56202.005400  & 325  & $-$ 0.000075  \\ 
        56179.233880  & 92  & 0.000387  & 56190.668720  & 209  & 0.000374  & 56202.103360  & 326  & 0.000149  \\
        56179.331530  & 93  & 0.000301  & 56190.766370  & 210  & 0.000291  & 56202.201190  & 327  & 0.000245  \\ 
        56179.428510  & 94  & $-$ 0.000451  & 56190.864230  & 211  & 0.000408  & 56202.298940  & 328  & 0.000268  \\ 
        56179.527210  & 95  & 0.000520  & 56190.962000  & 212  & 0.000449  & 56202.396470  & 329  & 0.000060  \\ 
        56179.623860  & 96  & $-$ 0.000570  & 56191.059450  & 213  & 0.000167  & 56202.493950  & 330  & $-$ 0.000192  \\ 
        56179.722860  & 97  & 0.000702  & 56191.157180  & 214  & 0.000160  & 56202.591990  & 331  & 0.000110  \\ 
        56179.820160  & 98  & 0.000263  & 56191.254620  & 215  & $-$ 0.000130  & 56202.689860  & 332  & 0.000252  \\ 
        56179.918350  & 99  & 0.000723  & 56191.352920  & 216  & 0.000429  & 56202.786320  & 333  & $-$ 0.001022  \\ 
        56180.015740  & 100  & 0.000379  & 56191.450490  & 217  & 0.000272  & 56202.885650  & 334  & 0.000573  \\ 
        56180.113300  & 101  & 0.000206  & 56191.547920  & 218  & $-$ 0.000030  & 56202.982810  & 335  & $-$ 0.000003  \\ 
        56180.210870  & 102  & 0.000038  & 56191.645170  & 219  & $-$ 0.000518  & 56203.078690  & 336  & $-$ 0.001854  \\ 
        56180.306740  & 103  & $-$ 0.001828  & 56191.743720  & 220  & 0.000298  & 56203.178700  & 337  & 0.000419  \\ 
        56180.406420  & 104  & 0.000120  & 56191.840570  & 221  & $-$ 0.000588  & 56203.276280  & 338  & 0.000264  \\ 
        56180.504280  & 105  & 0.000248  & 56191.937960  & 222  & $-$ 0.000929  & 56203.373270  & 339  & $-$ 0.000481  \\ 
        56180.601890  & 106  & 0.000127  & 56192.036740  & 223  & 0.000114  & 56203.471890  & 340  & 0.000414  \\ 
        56180.699160  & 107  & $-$ 0.000344  & 56192.134440  & 224  & 0.000079  & 56203.567240  & 341  & $-$ 0.001976  \\ 
        56180.795700  & 108  & $-$ 0.001533  & 56192.232960  & 225  & 0.000872  & 56203.667720  & 342  & 0.000768  \\ 
        56180.895050  & 109  & 0.000080  & 56192.329310  & 226  & $-$ 0.000514  & 56203.764400  & 343  & $-$ 0.000277  \\ 
        56180.992300  & 110  & $-$ 0.000399  & 56192.427230  & 227  & $-$ 0.000330  & 56203.862380  & 344  & $-$ 0.000035  \\ 
        56181.089960  & 111  & $-$ 0.000480  & 56192.525790  & 228  & 0.000494  & 56203.960520  & 345  & 0.000374  \\ 
        56181.188630  & 112  & 0.000464  & 56192.623420  & 229  & 0.000396  & 56204.058200  & 346  & 0.000315  \\ 
        56181.384160  & 114  & 0.000520  & 56192.719440  & 230  & $-$ 0.001320  & 56204.155130  & 347  & $-$ 0.000484  \\ 
        56181.481690  & 115  & 0.000322  & 56192.818760  & 231  & 0.000262  & 56204.253790  & 348  & 0.000439  \\ 

   \enddata
\tablecomments{$T_{max}$ is the observed times light maxima of Q14.3. E: Cycle number. $O - C$ is in days. E and $O - C$ are based on the ephemeris formula: $T_{max}$=$T_{0}+P \times E$=2456170.241912(0)+0.097734(1) $\times$ $E$.}
\end{deluxetable}

\begin{deluxetable}{ccrccrccr}
\tabletypesize{\small}
\tablewidth{0pc}
\tablenum{4}
\tablecaption{351 times of Maximum Light and $O - C$ Values of Q15.3.
\label{tab:max 153}} 
\tablehead{
\colhead{$BJD$}   &
\colhead{$E$}  &
\colhead{$O-C$ }      &
\colhead{$BJD$}            &
\colhead{$E$} &
\colhead{$O-C$ }      &
\colhead{$BJD$}            &
\colhead{$E$} &
\colhead{$O-C$} \\
\colhead{(2400000+)}   &
\colhead{}  &
\colhead{(day)}      &
\colhead{(2400000+)}            &
\colhead{} &
\colhead{ (day)}      &
\colhead{(2400000+)}            &
\colhead{} &
\colhead{(day)}   
}
\startdata
        56269.247220  & 1013  & $-$ 0.000625  & 56280.877690  & 1132  & $-$ 0.000758  & 56292.605340  & 1252  & $-$ 0.000334  \\ 
        56269.344600  & 1014  & $-$ 0.000279  & 56280.974730  & 1133  & $-$ 0.000064  & 56292.702980  & 1253  & $-$ 0.000240  \\ 
        56269.442220  & 1015  & $-$ 0.000161  & 56281.073020  & 1134  & $-$ 0.000620  & 56292.800910  & 1254  & $-$ 0.000436  \\ 
        56269.540260  & 1016  & $-$ 0.000464  & 56281.170630  & 1135  & $-$ 0.000496  & 56292.898380  & 1255  & $-$ 0.000172  \\ 
        56269.637800  & 1017  & $-$ 0.000275  & 56281.268270  & 1136  & $-$ 0.000402  & 56292.996080  & 1256  & $-$ 0.000138  \\ 
        56269.735600  & 1018  & $-$ 0.000335  & 56281.365640  & 1137  & $-$ 0.000038  & 56293.094110  & 1257  & $-$ 0.000434  \\ 
        56269.833090  & 1019  & $-$ 0.000091  & 56281.463680  & 1138  & $-$ 0.000344  & 56293.191810  & 1258  & $-$ 0.000400  \\ 
        56269.931190  & 1020  & $-$ 0.000464  & 56281.561340  & 1139  & $-$ 0.000271  & 56293.289020  & 1259  & 0.000124  \\ 
        56270.026710  & 1021  & 0.001753  & 56281.657010  & 1140  & 0.001793  & 56293.386910  & 1260  & $-$ 0.000032  \\ 
        56270.126490  & 1022  & $-$ 0.000293  & 56281.756560  & 1141  & $-$ 0.000023  & 56293.484200  & 1261  & 0.000412  \\ 
        56270.223990  & 1023  & $-$ 0.000059  & 56281.854750  & 1142  & $-$ 0.000479  & 56293.582370  & 1262  & $-$ 0.000024  \\ 
        56270.321960  & 1024  & $-$ 0.000295  & 56281.952060  & 1143  & $-$ 0.000055  & 56293.679450  & 1263  & 0.000630  \\ 
        56270.419330  & 1025  & 0.000069  & 56282.049780  & 1144  & $-$ 0.000041  & 56293.778280  & 1264  & $-$ 0.000466  \\ 
        56270.517320  & 1026  & $-$ 0.000187  & 56282.148230  & 1145  & $-$ 0.000757  & 56293.876100  & 1265  & $-$ 0.000552  \\ 
        56270.615050  & 1027  & $-$ 0.000183  & 56282.245210  & 1146  & $-$ 0.000003  & 56293.973760  & 1266  & $-$ 0.000479  \\ 
        56270.712930  & 1028  & $-$ 0.000329  & 56282.343250  & 1147  & $-$ 0.000309  & 56294.071190  & 1267  & $-$ 0.000175  \\ 
        56270.809880  & 1029  & 0.000455  & 56282.441320  & 1148  & $-$ 0.000645  & 56294.169130  & 1268  & $-$ 0.000381  \\ 
        56270.908090  & 1030  & $-$ 0.000021  & 56282.538800  & 1149  & $-$ 0.000391  & 56294.267110  & 1269  & $-$ 0.000627  \\
        56271.005880  & 1031  & $-$ 0.000077  & 56282.636340  & 1150  & $-$ 0.000197  & 56294.364320  & 1270  & $-$ 0.000103  \\
        56271.103900  & 1032  & $-$ 0.000363  & 56282.734100  & 1151  & $-$ 0.000223  & 56294.462540  & 1271  & $-$ 0.000589  \\ 
        56271.201530  & 1033  & $-$ 0.000260  & 56282.831650  & 1152  & $-$ 0.000039  & 56294.560070  & 1272  & $-$ 0.000385  \\ 
        56271.299310  & 1034  & $-$ 0.000306  & 56282.929800  & 1153  & $-$ 0.000455  & 56294.657310  & 1273  & 0.000109  \\ 
        56271.396650  & 1035  & 0.000088  & 56283.026340  & 1154  & 0.000739  & 56294.755720  & 1274  & $-$ 0.000567  \\ 
        56271.495050  & 1036  & $-$ 0.000578  & 56283.125400  & 1155  & $-$ 0.000587  & 56294.852860  & 1275  & 0.000027  \\ 
        56271.592500  & 1037  & $-$ 0.000294  & 56283.221640  & 1156  & 0.000907  & 56294.950880  & 1276  & $-$ 0.000259  \\ 
        56271.689900  & 1038  & 0.000040  & 56283.319670  & 1157  & 0.000611  & 56295.048320  & 1277  & 0.000035  \\ 
        56271.788110  & 1039  & $-$ 0.000436  & 56283.418560  & 1158  & $-$ 0.000545  & 56295.146050  & 1278  & 0.000039  \\ 
        56271.885550  & 1040  & $-$ 0.000142  & 56283.613210  & 1160  & 0.000272  & 56295.244090  & 1279  & $-$ 0.000267  \\ 
        56271.983470  & 1041  & $-$ 0.000328  & 56283.711590  & 1161  & $-$ 0.000374  & 56295.342260  & 1280  & $-$ 0.000703  \\ 
        56272.081340  & 1042  & $-$ 0.000464  & 56283.809150  & 1162  & $-$ 0.000200  & 56295.439120  & 1281  & 0.000171  \\
        56272.178820  & 1043  & $-$ 0.000210  & 56283.907030  & 1163  & $-$ 0.000346  & 56295.537060  & 1282  & $-$ 0.000035  \\
        56272.276530  & 1044  & $-$ 0.000186  & 56284.004850  & 1164  & $-$ 0.000432  & 56295.634240  & 1283  & 0.000519  \\ 
        56272.374310  & 1045  & $-$ 0.000232  & 56284.102380  & 1165  & $-$ 0.000228  & 56295.729570  & 1284  & 0.002923  \\ 
        56272.471890  & 1046  & $-$ 0.000078  & 56284.199830  & 1166  & 0.000056  & 56295.829430  & 1285  & 0.000797  \\ 
        56272.569900  & 1047  & $-$ 0.000354  & 56284.297950  & 1167  & $-$ 0.000330  & 56295.926660  & 1286  & 0.001301  \\ 
        56272.667660  & 1048  & $-$ 0.000380  & 56284.395300  & 1168  & 0.000054  & 56296.026600  & 1287  & $-$ 0.000906  \\ 
        56272.765320  & 1049  & $-$ 0.000306  & 56284.493620  & 1169  & $-$ 0.000532  & 56296.123170  & 1288  & 0.000258  \\ 
        56272.863000  & 1050  & $-$ 0.000252  & 56284.591050  & 1170  & $-$ 0.000228  & 56296.221370  & 1289  & $-$ 0.000208  \\ 
        56272.960750  & 1051  & $-$ 0.000268  & 56284.687600  & 1171  & 0.000956  & 56296.317860  & 1290  & 0.001036  \\ 
        56273.058280  & 1052  & $-$ 0.000064  & 56284.785250  & 1172  & 0.001040  & 56296.414040  & 1291  & 0.002590  \\ 
        56273.155470  & 1053  & 0.000480  & 56284.884340  & 1173  & $-$ 0.000316  & 56296.511000  & 1292  & 0.003364  \\ 
        56273.253740  & 1054  & $-$ 0.000057  & 56284.981990  & 1174  & $-$ 0.000232  & 56296.611910  & 1293  & 0.000188  \\ 
        56273.351600  & 1055  & $-$ 0.000183  & 56285.077820  & 1175  & 0.001672  & 56296.709150  & 1294  & 0.000682  \\ 
        56273.448980  & 1056  & 0.000171  & 56285.177470  & 1176  & $-$ 0.000244  & 56296.807790  & 1295  & $-$ 0.000224  \\ 
        56273.547220  & 1057  & $-$ 0.000335  & 56285.275130  & 1177  & $-$ 0.000170  & 56296.905310  & 1296  & $-$ 0.000010  \\ 
        56273.644420  & 1058  & 0.000199  & 56285.372050  & 1178  & 0.000644  & 56297.002910  & 1297  & 0.000124  \\ 
        56273.741280  & 1059  & 0.001073  & 56285.467800  & 1179  & 0.002628  & 56297.100650  & 1298  & 0.000118  \\ 
        56273.938520  & 1061  & $-$ 0.000699  & 56285.565450  & 1180  & 0.002712  & 56297.197940  & 1299  & 0.000562  \\ 
        56274.035230  & 1062  & 0.000325  & 56285.665800  & 1181  & 0.000095  & 56297.296390  & 1300  & $-$ 0.000154  \\ 
        56274.133570  & 1063  & $-$ 0.000281  & 56285.861330  & 1183  & 0.000033  & 56297.392410  & 1301  & 0.001560  \\ 
        56274.231150  & 1064  & $-$ 0.000127  & 56285.959600  & 1184  & $-$ 0.000503  & 56297.491260  & 1302  & 0.000444  \\ 
        56274.327110  & 1065  & 0.001647  & 56286.056500  & 1185  & 0.000331  & 56297.589490  & 1303  & $-$ 0.000052  \\ 
        56274.426320  & 1066  & 0.000171  & 56286.155120  & 1186  & $-$ 0.000555  & 56297.785160  & 1305  & $-$ 0.000254  \\ 
        56274.524000  & 1067  & 0.000225  & 56286.252810  & 1187  & $-$ 0.000511  & 56297.882660  & 1306  & $-$ 0.000020  \\ 
        56274.622160  & 1068  & $-$ 0.000201  & 56286.350580  & 1188  & $-$ 0.000547  & 56297.980570  & 1307  & $-$ 0.000196  \\ 
        56274.717460  & 1069  & 0.002233  & 56286.448300  & 1189  & $-$ 0.000533  & 56298.078420  & 1308  & $-$ 0.000313  \\ 
        56274.817810  & 1070  & $-$ 0.000383  & 56286.545780  & 1190  & $-$ 0.000279  & 56298.175800  & 1309  & 0.000041  \\ 
        56274.915010  & 1071  & 0.000151  & 56286.643040  & 1191  & 0.000195  & 56298.273730  & 1310  & $-$ 0.000155  \\ 
        56275.013020  & 1072  & $-$ 0.000125  & 56286.741250  & 1192  & $-$ 0.000281  & 56298.372080  & 1311  & $-$ 0.000771  \\ 
        56275.111010  & 1073  & $-$ 0.000381  & 56286.838980  & 1193  & $-$ 0.000277  & 56298.469360  & 1312  & $-$ 0.000317  \\ 
        56275.208390  & 1074  & $-$ 0.000027  & 56286.936360  & 1194  & 0.000077  & 56298.566870  & 1313  & $-$ 0.000093  \\ 
        56275.306490  & 1075  & $-$ 0.000394  & 56287.034590  & 1195  & $-$ 0.000419  & 56298.665140  & 1314  & $-$ 0.000629  \\
        56275.404250  & 1076  & $-$ 0.000420  & 56287.131680  & 1196  & 0.000225  & 56298.761470  & 1315  & 0.000775  \\ 
        56275.501750  & 1077  & $-$ 0.000186  & 56287.229660  & 1197  & $-$ 0.000021  & 56298.860050  & 1316  & $-$ 0.000071  \\ 
        56275.599390  & 1078  & $-$ 0.000092  & 56287.327810  & 1198  & $-$ 0.000437  & 56298.958050  & 1317  & $-$ 0.000337  \\ 
        56275.697110  & 1079  & $-$ 0.000078  & 56287.424240  & 1199  & 0.000867  & 56299.055550  & 1318  & $-$ 0.000103  \\ 
        56275.794350  & 1080  & 0.000416  & 56287.522850  & 1200  & $-$ 0.000009  & 56299.153290  & 1319  & $-$ 0.000109  \\ 
        56275.892650  & 1081  & $-$ 0.000150  & 56287.619450  & 1201  & 0.001125  & 56299.251170  & 1320  & $-$ 0.000255  \\ 
        56275.990500  & 1082  & $-$ 0.000266  & 56287.718830  & 1202  & $-$ 0.000522  & 56299.348430  & 1321  & 0.000219  \\
        56276.087580  & 1083  & 0.000388  & 56287.816400  & 1203  & $-$ 0.000358  & 56299.446700  & 1322  & $-$ 0.000317  \\ 
        56276.185870  & 1084  & $-$ 0.000168  & 56287.913600  & 1204  & 0.000176  & 56299.544730  & 1323  & $-$ 0.000613  \\
        56276.283210  & 1085  & 0.000226  & 56288.011850  & 1205  & $-$ 0.000340  & 56299.641680  & 1324  & 0.000171  \\
        56276.381720  & 1086  & $-$ 0.000550  & 56288.109180  & 1206  & 0.000064  & 56299.739800  & 1325  & $-$ 0.000215  \\ 
        56276.478890  & 1087  & 0.000014  & 56288.204520  & 1207  & 0.002458  & 56299.837890  & 1326  & $-$ 0.000571  \\ 
        56276.576340  & 1088  & 0.000298  & 56288.305390  & 1208  & $-$ 0.000678  & 56299.935590  & 1327  & $-$ 0.000537  \\
        56276.674680  & 1089  & $-$ 0.000308  & 56288.402210  & 1209  & 0.000236  & 56300.033080  & 1328  & $-$ 0.000293  \\ 
        56276.772290  & 1090  & $-$ 0.000184  & 56288.500550  & 1210  & $-$ 0.000370  & 56300.130640  & 1329  & $-$ 0.000119  \\ 
        56276.868770  & 1091  & 0.001070  & 56288.596770  & 1211  & 0.001144  & 56300.228430  & 1330  & $-$ 0.000176  \\ 
        56276.967910  & 1092  & $-$ 0.000336  & 56288.695490  & 1212  & 0.000158  & 56300.326330  & 1331  & $-$ 0.000342  \\ 
        56277.065380  & 1093  & $-$ 0.000072  & 56288.793390  & 1213  & $-$ 0.000008  & 56300.424350  & 1332  & $-$ 0.000628  \\
        56277.163150  & 1094  & $-$ 0.000108  & 56288.891480  & 1214  & $-$ 0.000364  & 56300.521650  & 1333  & $-$ 0.000194  \\ 
        56277.261000  & 1095  & $-$ 0.000224  & 56288.989070  & 1215  & $-$ 0.000220  & 56300.617750  & 1334  & 0.001440  \\ 
        56277.358660  & 1096  & $-$ 0.000151  & 56289.086150  & 1216  & 0.000434  & 56300.717080  & 1335  & $-$ 0.000156  \\ 
        56277.456510  & 1097  & $-$ 0.000267  & 56289.184800  & 1217  & $-$ 0.000482  & 56300.814960  & 1336  & $-$ 0.000302  \\ 
        56277.554080  & 1098  & $-$ 0.000103  & 56289.282340  & 1218  & $-$ 0.000288  & 56300.912480  & 1337  & $-$ 0.000088  \\
        56277.652160  & 1099  & $-$ 0.000449  & 56289.380040  & 1219  & $-$ 0.000254  & 56301.010610  & 1338  & $-$ 0.000484  \\ 
        56277.749520  & 1100  & $-$ 0.000075  & 56289.477720  & 1220  & $-$ 0.000200  & 56301.108370  & 1339  & $-$ 0.000510  \\
        56277.945130  & 1102  & $-$ 0.000217  & 56289.673180  & 1222  & $-$ 0.000192  & 56301.303850  & 1341  & $-$ 0.000522  \\ 
        56278.043150  & 1103  & $-$ 0.000503  & 56289.771040  & 1223  & $-$ 0.000318  & 56301.401420  & 1342  & $-$ 0.000358  \\ 
        56278.140930  & 1104  & $-$ 0.000549  & 56289.868640  & 1224  & $-$ 0.000185  & 56301.498990  & 1343  & $-$ 0.000194  \\ 
        56278.238740  & 1105  & $-$ 0.000625  & 56289.965950  & 1225  & 0.000239  & 56301.597170  & 1344  & $-$ 0.000640  \\ 
        56278.336400  & 1106  & $-$ 0.000551  & 56290.064420  & 1226  & $-$ 0.000497  & 56301.694170  & 1345  & 0.000094  \\
        56278.431730  & 1107  & 0.001853  & 56290.160130  & 1227  & 0.001527  & 56301.792080  & 1346  & $-$ 0.000082  \\ 
        56278.531020  & 1108  & 0.000297  & 56290.259940  & 1228  & $-$ 0.000549  & 56301.889740  & 1347  & $-$ 0.000008  \\ 
        56278.629460  & 1109  & $-$ 0.000409  & 56290.357620  & 1229  & $-$ 0.000495  & 56301.987940  & 1348  & $-$ 0.000474  \\ 
        56278.726700  & 1110  & 0.000085  & 56290.455440  & 1230  & $-$ 0.000581  & 56302.085530  & 1349  & $-$ 0.000330  \\ 
        56278.824790  & 1111  & $-$ 0.000271  & 56290.553010  & 1231  & $-$ 0.000417  & 56302.183430  & 1350  & $-$ 0.000496  \\
        56278.921720  & 1112  & 0.000533  & 56290.650280  & 1232  & 0.000047  & 56302.280910  & 1351  & $-$ 0.000243  \\ 
        56279.020440  & 1113  & $-$ 0.000453  & 56290.748250  & 1233  & $-$ 0.000189  & 56302.378520  & 1352  & $-$ 0.000119  \\ 
        56279.118050  & 1114  & $-$ 0.000329  & 56290.846020  & 1234  & $-$ 0.000225  & 56302.475930  & 1353  & 0.000205  \\
        56279.215760  & 1115  & $-$ 0.000305  & 56290.943910  & 1235  & $-$ 0.000381  & 56302.574170  & 1354  & $-$ 0.000301  \\ 
        56279.313150  & 1116  & 0.000039  & 56291.041540  & 1236  & $-$ 0.000277  & 56302.672260  & 1355  & $-$ 0.000657  \\
        56279.409370  & 1117  & 0.001553  & 56291.139230  & 1237  & $-$ 0.000233  & 56302.768850  & 1356  & 0.000487  \\ 
        56279.508740  & 1118  & $-$ 0.000084  & 56291.237000  & 1238  & $-$ 0.000269  & 56302.867500  & 1357  & $-$ 0.000429  \\ 
        56279.606250  & 1119  & 0.000140  & 56291.334790  & 1239  & $-$ 0.000325  & 56302.965130  & 1358  & $-$ 0.000325  \\
        56279.703700  & 1120  & 0.000424  & 56291.432170  & 1240  & 0.000029  & 56303.062770  & 1359  & $-$ 0.000231  \\ 
        56279.898990  & 1122  & 0.000602  & 56291.530090  & 1241  & $-$ 0.000157  & 56303.160200  & 1360  & 0.000073  \\ 
        56279.996930  & 1123  & 0.000396  & 56291.626750  & 1242  & 0.000917  & 56303.256340  & 1361  & 0.001667  \\ 
        56280.095200  & 1124  & $-$ 0.000140  & 56291.823050  & 1244  & 0.000085  & 56303.356080  & 1362  & $-$ 0.000339  \\ 
        56280.192880  & 1125  & $-$ 0.000086  & 56291.920980  & 1245  & $-$ 0.000112  & 56303.453260  & 1363  & 0.000215  \\ 
        56280.286690  & 1126  & 0.003838  & 56292.018040  & 1246  & 0.000562  & 56303.551610  & 1364  & $-$ 0.000401  \\ 
        56280.388840  & 1127  & $-$ 0.000578  & 56292.114970  & 1247  & 0.001366  & 56303.649390  & 1365  & $-$ 0.000447  \\ 
        56280.483200  & 1128  & 0.002796  & 56292.214200  & 1248  & $-$ 0.000130  & 56303.746520  & 1366  & 0.000157  \\
        56280.584210  & 1129  & $-$ 0.000480  & 56292.311990  & 1249  & $-$ 0.000186  & 56303.844710  & 1367  & $-$ 0.000299  \\ 
        56280.680730  & 1130  & 0.000734  & 56292.409870  & 1250  & $-$ 0.000332  & 56303.942220  & 1368  & $-$ 0.000075  \\ 
        56280.779730  & 1131  & $-$ 0.000532  & 56292.507010  & 1251  & 0.000262  & 56304.040060  & 1369  & $-$ 0.000181  \\ 
        
   \enddata
\tablecomments{$T_{max}$ is the observed times light maxima of Q15.3. E: Cycle number. $O - C$ is in days. E and $O - C$ are based on the ephemeris formula: $T_{max}$=$T_{0}+P \times E$=2456170.241912(0)+0.097734(1)$\times$ $E$.}
\end{deluxetable}

\begin{deluxetable}{ccrccrccr}
\tabletypesize{\small}
\tablewidth{0pc}
\tablenum{5}
\tablecaption{322 times of Maximum Light and $O - C$ Values of Q16.3.
\label{tab:max 163}} 
\tablehead{
\colhead{$BJD$}   &
\colhead{$E$}  &
\colhead{$O-C$ }      &
\colhead{$BJD$}            &
\colhead{$E$} &
\colhead{$O-C$ }      &
\colhead{$BJD$}            &
\colhead{$E$} &
\colhead{$O-C$} \\
\colhead{(2400000+)}   &
\colhead{}  &
\colhead{(day)}      &
\colhead{(2400000+)}            &
\colhead{} &
\colhead{ (day)}      &
\colhead{(2400000+)}            &
\colhead{} &
\colhead{(day)}   
}
\startdata
        56359.162380  & 1933  & $-$ 0.000284  & 56369.717590  & 2041  & $-$ 0.000250  & 56380.465280  & 2151  & 0.002771  \\
        56359.259800  & 1934  & 0.000030  & 56369.814310  & 2042  & 0.000764  & 56380.564130  & 2152  & 0.001655  \\ 
        56359.358020  & 1935  & $-$ 0.000457  & 56369.912870  & 2043  & $-$ 0.000063  & 56380.664030  & 2153  & $-$ 0.000511  \\ 
        56359.455730  & 1936  & $-$ 0.000433  & 56370.010790  & 2044  & $-$ 0.000249  & 56380.761410  & 2154  & $-$ 0.000157  \\ 
        56359.553590  & 1937  & $-$ 0.000559  & 56370.108430  & 2045  & $-$ 0.000155  & 56380.858970  & 2155  & 0.000016  \\ 
        56359.650790  & 1938  & $-$ 0.000025  & 56370.206570  & 2046  & $-$ 0.000561  & 56380.956700  & 2156  & 0.000020  \\ 
        56359.748350  & 1939  & 0.000148  & 56370.303269  & 2047  & 0.000473  & 56381.055070  & 2157  & $-$ 0.000616  \\ 
        56359.846140  & 1940  & 0.000092  & 56370.401730  & 2048  & $-$ 0.000254  & 56381.151830  & 2158  & 0.000358  \\ 
        56359.943890  & 1941  & 0.000076  & 56370.499680  & 2049  & $-$ 0.000470  & 56381.250210  & 2159  & $-$ 0.000289  \\ 
        56360.042020  & 1942  & $-$ 0.000320  & 56370.596596  & 2050  & 0.000348  & 56381.347850  & 2160  & $-$ 0.000195  \\ 
        56360.139360  & 1943  & 0.000073  & 56370.694910  & 2051  & $-$ 0.000233  & 56381.445770  & 2161  & $-$ 0.000381  \\ 
        56360.236871  & 1944  & 0.000296  & 56370.792510  & 2052  & $-$ 0.000099  & 56381.543410  & 2162  & $-$ 0.000288  \\ 
        56360.334990  & 1945  & $-$ 0.000089  & 56370.890860  & 2053  & $-$ 0.000715  & 56381.641120  & 2163  & $-$ 0.000264  \\ 
        56360.433060  & 1946  & $-$ 0.000425  & 56370.988130  & 2054  & $-$ 0.000251  & 56381.738450  & 2164  & 0.000140  \\ 
        56360.530830  & 1947  & $-$ 0.000462  & 56371.085640  & 2055  & $-$ 0.000028  & 56381.934710  & 2166  & $-$ 0.000653  \\ 
        56360.627840  & 1948  & 0.000262  & 56371.183550  & 2056  & $-$ 0.000204  & 56382.030950  & 2167  & 0.000841  \\
        56360.725940  & 1949  & $-$ 0.000104  & 56371.280630  & 2057  & 0.000450  & 56382.129500  & 2168  & 0.000025  \\ 
        56360.823820  & 1950  & $-$ 0.000250  & 56371.378620  & 2058  & 0.000193  & 56382.227340  & 2169  & $-$ 0.000081  \\ 
        56360.921700  & 1951  & $-$ 0.000397  & 56371.475570  & 2059  & 0.000977  & 56382.325380  & 2170  & $-$ 0.000388  \\ 
        56361.019220  & 1952  & $-$ 0.000183  & 56371.574460  & 2060  & $-$ 0.000179  & 56382.420380  & 2171  & 0.002346  \\ 
        56361.116850  & 1953  & $-$ 0.000079  & 56371.672420  & 2061  & $-$ 0.000405  & 56382.520870  & 2172  & $-$ 0.000410  \\ 
        56361.214470  & 1954  & 0.000034  & 56371.769780  & 2062  & $-$ 0.000032  & 56382.617780  & 2173  & 0.000414  \\
        56361.312230  & 1955  & 0.000008  & 56371.867980  & 2063  & $-$ 0.000498  & 56382.716120  & 2174  & $-$ 0.000193  \\ 
        56361.410080  & 1956  & $-$ 0.000108  & 56371.965300  & 2064  & $-$ 0.000084  & 56382.814270  & 2175  & $-$ 0.000609  \\ 
        56361.507840  & 1957  & $-$ 0.000134  & 56372.062360  & 2065  & 0.000590  & 56382.911680  & 2176  & $-$ 0.000285  \\ 
        56361.605440  & 1958  & $-$ 0.000001  & 56372.159900  & 2066  & 0.000783  & 56383.009130  & 2177  & $-$ 0.000001  \\
        56361.702870  & 1959  & 0.000303  & 56372.258650  & 2067  & $-$ 0.000233  & 56383.105760  & 2178  & 0.001102  \\ 
        56361.800750  & 1960  & 0.000157  & 56372.355490  & 2068  & 0.000661  & 56383.204740  & 2179  & $-$ 0.000144  \\ 
        56361.898570  & 1961  & 0.000071  & 56372.454240  & 2069  & $-$ 0.000355  & 56383.301010  & 2180  & 0.001320  \\ 
        56361.996480  & 1962  & $-$ 0.000106  & 56372.551120  & 2070  & 0.000498  & 56383.400210  & 2181  & $-$ 0.000146  \\ 
        56362.094180  & 1963  & $-$ 0.000072  & 56372.649430  & 2071  & $-$ 0.000078  & 56383.497060  & 2182  & 0.000737  \\ 
        56362.191930  & 1964  & $-$ 0.000088  & 56372.747700  & 2072  & $-$ 0.000614  & 56383.595710  & 2183  & $-$ 0.000179  \\ 
        56362.289900  & 1965  & $-$ 0.000324  & 56372.844680  & 2073  & 0.000140  & 56383.693530  & 2184  & $-$ 0.000265  \\ 
        56362.387470  & 1966  & $-$ 0.000161  & 56372.942540  & 2074  & 0.000013  & 56383.791270  & 2185  & $-$ 0.000271  \\ 
        56362.484990  & 1967  & 0.000053  & 56373.040660  & 2075  & $-$ 0.000373  & 56383.889080  & 2186  & $-$ 0.000348  \\ 
        56362.582540  & 1968  & 0.000237  & 56373.137930  & 2076  & 0.000091  & 56383.986200  & 2187  & 0.000266  \\ 
        56362.680990  & 1969  & $-$ 0.000479  & 56373.236100  & 2077  & $-$ 0.000345  & 56384.084340  & 2188  & $-$ 0.000140  \\ 
        56362.778290  & 1970  & $-$ 0.000046  & 56373.333630  & 2078  & $-$ 0.000142  & 56384.182220  & 2189  & $-$ 0.000287  \\ 
        56362.875180  & 1971  & 0.000798  & 56373.431490  & 2079  & $-$ 0.000268  & 56384.280250  & 2190  & $-$ 0.000583  \\ 
        56362.974260  & 1972  & $-$ 0.000548  & 56373.528860  & 2080  & 0.000096  & 56384.377650  & 2191  & $-$ 0.000249  \\ 
        56363.070610  & 1973  & 0.000836  & 56373.626509  & 2081  & 0.000181  & 56384.475830  & 2192  & $-$ 0.000695  \\ 
        56363.169520  & 1974  & $-$ 0.000341  & 56373.724830  & 2082  & $-$ 0.000407  & 56384.573190  & 2193  & $-$ 0.000322  \\ 
        56363.267200  & 1975  & $-$ 0.000287  & 56373.822560  & 2083  & $-$ 0.000403  & 56384.671350  & 2194  & $-$ 0.000748  \\ 
        56363.365161  & 1976  & $-$ 0.000514  & 56373.920060  & 2084  & $-$ 0.000169  & 56384.866320  & 2196  & $-$ 0.000250  \\ 
        56363.462480  & 1977  & $-$ 0.000099  & 56374.018050  & 2085  & $-$ 0.000426  & 56384.963520  & 2197  & 0.000283  \\ 
        56363.560410  & 1978  & $-$ 0.000296  & 56374.115360  & 2086  & $-$ 0.000002  & 56385.061650  & 2198  & $-$ 0.000113  \\ 
        56363.658330  & 1979  & $-$ 0.000482  & 56374.213540  & 2087  & $-$ 0.000448  & 56385.156790  & 2199  & 0.002481  \\ 
        56363.755120  & 1980  & 0.000462  & 56374.310950  & 2088  & $-$ 0.000124  & 56385.257060  & 2200  & $-$ 0.000055  \\ 
        56363.853410  & 1981  & $-$ 0.000095  & 56374.408330  & 2089  & 0.000229  & 56385.355300  & 2201  & $-$ 0.000562  \\ 
        56363.951160  & 1982  & $-$ 0.000111  & 56374.506440  & 2090  & $-$ 0.000147  & 56385.451510  & 2202  & 0.000962  \\ 
        56364.048870  & 1983  & $-$ 0.000087  & 56374.604460  & 2091  & $-$ 0.000433  & 56385.549890  & 2203  & 0.000316  \\ 
        56364.146780  & 1984  & $-$ 0.000263  & 56374.701770  & 2092  & $-$ 0.000009  & 56385.648100  & 2204  & $-$ 0.000160  \\ 
        56364.244430  & 1985  & $-$ 0.000180  & 56374.798750  & 2093  & 0.000744  & 56385.745810  & 2205  & $-$ 0.000137  \\ 
        56364.342300  & 1986  & $-$ 0.000316  & 56374.897370  & 2094  & $-$ 0.000142  & 56385.843980  & 2206  & $-$ 0.000573  \\ 
        56364.440060  & 1987  & $-$ 0.000342  & 56374.995260  & 2095  & $-$ 0.000298  & 56385.941550  & 2207  & $-$ 0.000409  \\ 
        56364.537380  & 1988  & 0.000072  & 56375.093070  & 2096  & $-$ 0.000374  & 56386.039360  & 2208  & $-$ 0.000485  \\ 
        56364.634870  & 1989  & 0.000315  & 56375.190460  & 2097  & $-$ 0.000031  & 56386.136510  & 2209  & 0.000098  \\ 
        56364.732730  & 1990  & 0.000189  & 56375.288140  & 2098  & 0.000023  & 56386.234200  & 2210  & 0.000142  \\ 
        56364.830670  & 1991  & $-$ 0.000017  & 56375.386270  & 2099  & $-$ 0.000373  & 56386.331910  & 2211  & 0.000166  \\ 
        56364.927800  & 1992  & 0.000587  & 56375.483470  & 2100  & 0.000161  & 56386.430430  & 2212  & $-$ 0.000620  \\ 
        56365.025040  & 1993  & 0.001080  & 56375.581020  & 2101  & 0.000344  & 56386.527780  & 2213  & $-$ 0.000237  \\ 
        56365.123450  & 1994  & 0.000404  & 56375.679620  & 2102  & $-$ 0.000522  & 56386.625160  & 2214  & 0.000117  \\ 
        56365.221480  & 1995  & 0.000108  & 56375.777300  & 2103  & $-$ 0.000468  & 56386.723150  & 2215  & $-$ 0.000139  \\ 
        56365.319290  & 1996  & 0.000032  & 56375.874800  & 2104  & $-$ 0.000234  & 56386.820590  & 2216  & 0.000154  \\ 
        56365.417230  & 1997  & $-$ 0.000175  & 56375.972440  & 2105  & $-$ 0.000141  & 56386.918480  & 2217  & $-$ 0.000002  \\ 
        56365.515290  & 1998  & $-$ 0.000501  & 56376.069990  & 2106  & 0.000043  & 56387.015980  & 2218  & 0.000232  \\
        56365.612840  & 1999  & $-$ 0.000317  & 56376.167090  & 2107  & 0.000677  & 56387.113720  & 2219  & 0.000226  \\ 
        56365.710560  & 2000  & $-$ 0.000303  & 56376.265470  & 2108  & 0.000031  & 56387.211480  & 2220  & 0.000199  \\ 
        56365.808050  & 2001  & $-$ 0.000060  & 56376.363830  & 2109  & $-$ 0.000596  & 56387.309030  & 2221  & 0.000383  \\ 
        56365.905270  & 2002  & 0.000454  & 56376.461400  & 2110  & $-$ 0.000432  & 56387.407510  & 2222  & $-$ 0.000363  \\ 
        56366.003400  & 2003  & 0.000058  & 56376.558470  & 2111  & 0.000232  & 56387.504570  & 2223  & 0.000311  \\ 
        56366.101190  & 2004  & 0.000002  & 56376.656860  & 2112  & $-$ 0.000425  & 56387.602880  & 2224  & $-$ 0.000266  \\ 
        56366.198400  & 2005  & 0.000525  & 56376.754540  & 2113  & $-$ 0.000371  & 56387.700910  & 2225  & $-$ 0.000562  \\ 
        56366.296590  & 2006  & 0.000069  & 56376.852270  & 2114  & $-$ 0.000367  & 56387.798680  & 2226  & $-$ 0.000598  \\ 
        56366.394330  & 2007  & 0.000063  & 56376.950080  & 2115  & $-$ 0.000443  & 56387.896150  & 2227  & $-$ 0.000334  \\ 
        56366.492380  & 2008  & $-$ 0.000254  & 56377.047480  & 2116  & $-$ 0.000110  & 56387.993850  & 2228  & $-$ 0.000301  \\ 
        56366.590170  & 2009  & $-$ 0.000310  & 56377.145270  & 2117  & $-$ 0.000166  & 56388.091580  & 2229  & $-$ 0.000297  \\
        56366.687580  & 2010  & 0.000014  & 56377.243080  & 2118  & $-$ 0.000242  & 56388.188280  & 2230  & 0.000737  \\ 
        56366.785600  & 2011  & $-$ 0.000272  & 56377.340500  & 2119  & 0.000072  & 56388.286520  & 2231  & 0.000231  \\ 
        56366.883450  & 2012  & $-$ 0.000389  & 56377.438570  & 2120  & $-$ 0.000265  & 56388.384710  & 2232  & $-$ 0.000226  \\ 
        56366.981040  & 2013  & $-$ 0.000245  & 56377.536090  & 2121  & $-$ 0.000051  & 56388.481870  & 2233  & 0.000348  \\ 
        56367.077570  & 2014  & 0.000959  & 56377.633540  & 2122  & 0.000233  & 56388.580020  & 2234  & $-$ 0.000068  \\ 
        56367.176360  & 2015  & $-$ 0.000097  & 56377.731640  & 2123  & $-$ 0.000133  & 56388.677950  & 2235  & $-$ 0.000264  \\ 
        56367.274280  & 2016  & $-$ 0.000284  & 56377.829480  & 2124  & $-$ 0.000240  & 56388.775870  & 2236  & $-$ 0.000451  \\ 
        56367.371830  & 2017  & $-$ 0.000100  & 56377.927480  & 2125  & $-$ 0.000506  & 56388.873480  & 2237  & $-$ 0.000327  \\ 
        56367.469090  & 2018  & 0.000374  & 56378.024940  & 2126  & $-$ 0.000232  & 56388.971130  & 2238  & $-$ 0.000243  \\ 
        56367.567430  & 2019  & $-$ 0.000232  & 56378.122640  & 2127  & $-$ 0.000198  & 56389.067460  & 2239  & 0.001160  \\ 
        56367.665390  & 2020  & $-$ 0.000459  & 56378.220520  & 2128  & $-$ 0.000345  & 56389.166750  & 2240  & $-$ 0.000396  \\ 
        56367.762730  & 2021  & $-$ 0.000065  & 56378.317870  & 2129  & 0.000039  & 56389.264590  & 2241  & $-$ 0.000502  \\ 
        56367.860860  & 2022  & $-$ 0.000461  & 56378.415330  & 2130  & 0.000313  & 56389.361840  & 2242  & $-$ 0.000018  \\ 
        56367.957990  & 2023  & 0.000143  & 56378.513730  & 2131  & $-$ 0.000353  & 56389.459630  & 2243  & $-$ 0.000075  \\ 
        56368.055900  & 2024  & $-$ 0.000034  & 56378.611170  & 2132  & $-$ 0.000060  & 56389.557060  & 2244  & 0.000229  \\ 
        56368.153380  & 2025  & 0.000220  & 56378.707970  & 2133  & 0.000874  & 56389.655270  & 2245  & $-$ 0.000247  \\ 
        56368.251290  & 2026  & 0.000044  & 56378.807000  & 2134  & $-$ 0.000422  & 56389.753070  & 2246  & $-$ 0.000313  \\ 
        56368.349190  & 2027  & $-$ 0.000122  & 56378.904520  & 2135  & $-$ 0.000209  & 56389.850360  & 2247  & 0.000130  \\ 
        56368.447010  & 2028  & $-$ 0.000209  & 56379.002160  & 2136  & $-$ 0.000115  & 56389.948170  & 2248  & 0.000054  \\ 
        56368.544720  & 2029  & $-$ 0.000185  & 56379.100310  & 2137  & $-$ 0.000531  & 56390.044220  & 2249  & 0.001738  \\ 
        56368.632050  & 2030  & 0.010219  & 56379.197330  & 2138  & 0.000183  & 56390.144030  & 2250  & $-$ 0.000338  \\ 
        56368.740520  & 2031  & $-$ 0.000517  & 56379.295410  & 2139  & $-$ 0.000164  & 56390.241880  & 2251  & $-$ 0.000455  \\ 
        56368.838140  & 2032  & $-$ 0.000404  & 56379.393430  & 2140  & $-$ 0.000450  & 56390.338900  & 2252  & 0.000259  \\ 
        56368.936030  & 2033  & $-$ 0.000560  & 56379.588270  & 2142  & 0.000178  & 56390.437340  & 2253  & $-$ 0.000447  \\ 
        56369.033720  & 2034  & $-$ 0.000516  & 56379.685640  & 2143  & 0.000541  & 56390.534920  & 2254  & $-$ 0.000293  \\ 
        56369.131020  & 2035  & $-$ 0.000083  & 56379.784000  & 2144  & $-$ 0.000085  & 56390.632650  & 2255  & $-$ 0.000290  \\ 
        56369.228570  & 2036  & 0.000101  & 56379.880580  & 2145  & 0.001069  & 56390.730080  & 2256  & 0.000014  \\ 
        56369.326500  & 2037  & $-$ 0.000095  & 56379.979280  & 2146  & 0.000103  & 56390.828220  & 2257  & $-$ 0.000392  \\ 
        56369.423980  & 2038  & 0.000159  & 56380.075360  & 2147  & 0.001756  & 56390.925920  & 2258  & $-$ 0.000358  \\ 
        56369.522210  & 2039  & $-$ 0.000338  & 56380.174850  & 2148  & 0.000000  &  &  &  \\ 
        56369.618390  & 2040  & 0.001216  & 56380.369930  & 2150  & 0.000388  &  &  &  \\ 

   \enddata
\tablecomments{$T_{max}$ is the observed times light maxima of Q16.3. E: Cycle number. $O - C$ is in days. E and $O - C$ are based on the ephemeris formula: $T_{max}$=$T_{0}+P \times E$=2456170.241912(0)+0.097734(1)$\times$ $E$.}

\end{deluxetable}

\end{document}